\documentclass[pre,twocolumn,groupedaddress,showpacs,showkeys,amsmath,amssymb]{revtex4}

\usepackage{graphicx}
\usepackage{dcolumn}
\usepackage{bm}

\begin{document}

\title{Is negative-weight percolation compatible with SLE?}
\author{C. Norrenbrock}
\email{christoph.norrenbrock@uni-oldenburg.de}
\author{O. Melchert}
\email{oliver.melchert@uni-oldenburg.de}
\author{A. K. Hartmann}
\email{alexander.hartmann@uni-oldenburg.de}
\affiliation{
Institut f\"ur Physik, Universit\"at Oldenburg, 26111 Oldenburg, Germany
}
\date{\today}

\begin{abstract}
We study numerically the geometrical properties of minimally weighted paths that appear in the negative-weight percolation (NWP) model 
on two-dimensional lattices assuming a combination of periodic and free boundary conditions (BCs).
Each realization of the disorder consists of a random fraction $(1\!-\!\rho)$ of bonds with unit strength and a fraction $\rho$ of bond strengths drawn from a Gaussian distribution with zero mean and unit width.
For each such sample, the path is forced to span the lattice along the direction with the free BCs. The path and a set of negatively weighted loops form a ground state (GS).
A ground state on such a lattice can be determined performing a non-trivial transformation of the original graph and applying sophisticated matching algorithms.
Here we examine whether the geometrical properties of the paths are in accordance with predictions of Schramm-Loewner evolution (SLE). 
Measuring the fractal dimension and reviewing Schramm's left passage formula indicates that the paths cannot be described in terms of SLE.
\end{abstract} 

\maketitle

\section{Introduction \label{sect:introduction}}

The statistical properties of lattice-path models on graphs,
often equipped with quenched disorder, 
have experienced much attention during the  last decades.
Such paths can be as simple as boundaries of percolation
clusters \cite{stauffer1994} or domain walls in planar Ising systems
\cite{janke2005}.
Beyond this, they have proven to be useful in order to describe 
more complex line-like 
quantities as, e.g.,
linear polymers in disordered media 
\cite{kardar1987,derrida1990,grassberger1993,buldyrev2006}, 
vortices in  high $T_c$ superconductivity \cite{pfeiffer2002,pfeiffer2003},
cosmic strings in the early universe 
\cite{vachaspati1984,scherrer1986,hindmarsch1995},
and domain-wall excitations in disordered systems such as $2d$ 
spin glasses \cite{cieplak1994,melchert2007} and 
the $2d$ solid-on-solid model \cite{schwarz2009}. 
So as to analyze the statistical properties of these lattice path models, 
geometric observables and scaling concepts similar to those used in 
percolation theory \cite{stauffer1979,stauffer1994} or other string-bearing
models \cite{allega1990,austin1994} are often applicable. 
Recently,
the question which of these path models can be described in terms
of  Schramm-Loewner evolution (SLE) \cite{cardy2005,kager2004,bauer2006},
has attracted much attention. 

Basic problems like percolation or domain walls in Ising systems
can be treated numerically by standard approaches
\cite{stauffer1994, newman1999}. For more complex models, this
is not straight-forward. Nevertheless,
the precise computation of these paths can often be formulated in terms
of a combinatorial optimization problem and hence might allow for the
application of exact optimization algorithms developed in computer 
science \cite{schwartz1998,rieger2003}.

\begin{figure}[bt]
\centerline{
\includegraphics[width=1.\linewidth]{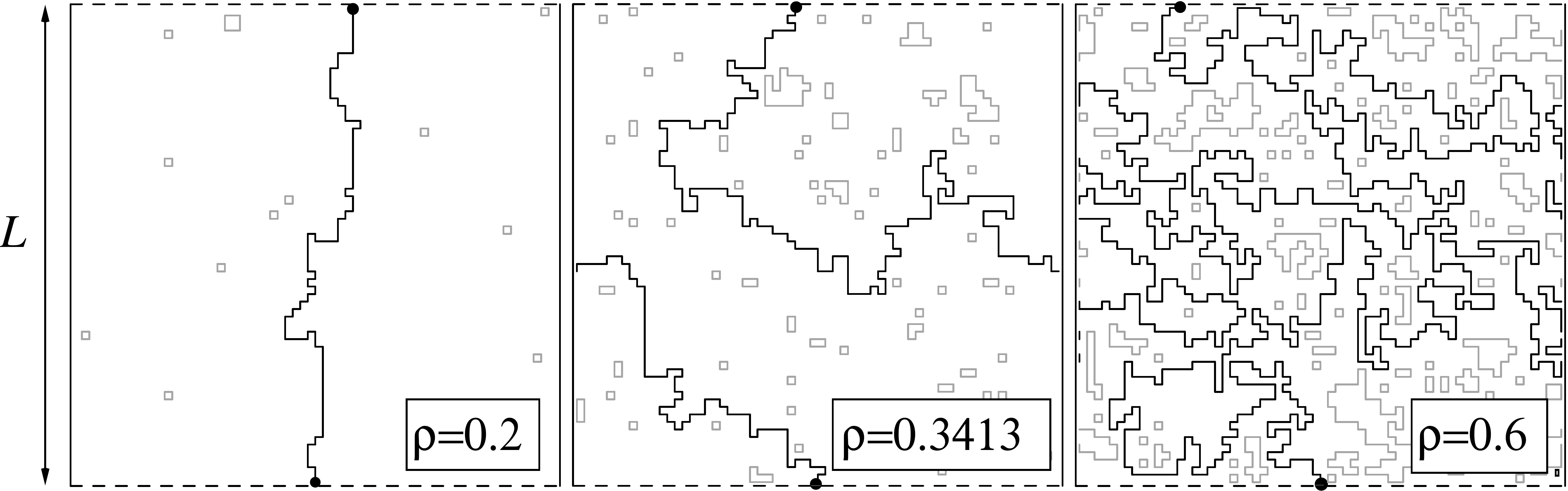}}
\caption{
Illustration of minimum-weight configurations consisting of loops (gray) and one path (black) in a regular $2d$ square lattice of side length $L\!=\!64$. 
The boundary conditions are periodic in horizontal (left-right)
direction and free in vertical direction.
The path is forced to connect the middle of the lower boundary to the upper boundary. 
The minimum-weight configurations minimize Eq.\ (\ref{eq:energy}).
For a small value of the disorder parameter $\rho\!=\!0.2$, the path 
crosses the lattice in a rather straight fashion, since the path 
weight is mainly determined by its length.
Increasing the order parameter to $\rho\!=\!0.3413$, the path 
length increases as well and the path takes advantage
of more negative-weighted edges. 
If the disorder parameter is increased to $\rho\!=\!0.6$, the loops and the path will become more dense.
\label{fig:alltogether}}
\end{figure}

In this article we study paths in the \emph{negative-weight percolation} 
(NWP) model \cite{melchert2008}.
The basis of this model is given by a weighted, undirected, $d$-dimensional lattice graph with side length $L$ where weights are assigned to the edges taken from a distribution that allows for values of either sign.
For a given realization of the disorder, a configuration consisting 
of one path, spanning the lattice along the direction of the free boundaries 
(the path extremities are forced to lie on the free boundaries),
and possibly a set of loops, i.e.\ closed paths, can be found such that the loops and the 
path do not intersect one another and the total sum of the weights assigned to the edges which 
build up this configuration attains an \emph{exact} minimum.
The fraction of negatively weighted edges can be controlled by a tunable disorder parameter $\rho$ that gives rise to qualitatively different ground states as depicted in Fig.\ \ref{fig:alltogether}.
As presented here, the NWP problem is a theoretical model of intrinsic interest.
As an example that makes use of the statement of the NWP problem, an agent can be imagined that travels on a graph.
While traversing an edge, the agent either needs to pay some resource (signified by a positive edge weight) or is able to harvest some resource (signified by a negative edge weight).
In order to gain as many resources as possible, the path/loops obtained in the context of the NWP problem serve as a guide to find optimal routes.

The problem of detecting the exact minimum-weight configuration for a given realization of the disorder becomes solvable through a mapping of the original graph to the minimum-weight perfect-matching optimization problem as outlined in Section \ref{sect:model}.
This mapping allows for applying exact polynomial-time-running algorithms, thus large instances can be solved.
Note that the same mapping and algorithms can be used in the context of finding ground-state spin configurations for the planar triangular random-bond Ising model, since this problem is equivalent to the NWP problem on a planar honeycomb lattice \cite{melchertThesis2009,oli2011}.

In a previous study it was shown that the NWP model features a disorder-driven, geometric phase transition \cite{melchert2008}.
Depending on the disorder parameter, two distinct phases can be identified: (i) a phase where the loops (if there are any) are small and the shape of the path is rather straight-lined, reflecting a self-affine scaling of the path length (cf.\ Fig.\ \ref{fig:alltogether}(a)) and (ii) a phase where large loops emerge and the extension of the horizontal projection of the path is $O(L)$, reflecting a self-similar scaling of the path length (cf.\ Fig.\ \ref{fig:alltogether}(c)).
In the limit of large system sizes, there is a particular value of the disorder parameter at which the path simultaneously spans the lattice along both directions for the first time (cf.\ Fig.\ \ref{fig:alltogether}(b)).
For two-dimensional (2D) lattice graphs, the respective disorder-driven phase transition was investigated using finite-size scaling analyses and characterized by a set of critical exponents.
It turned out that the exponents were universal in 2D and clearly distinct from those describing other percolation phenomena.
Regarding loops only, the influence of dilution on the critical properties of the 2D NWP phenomenon was investigated in a subsequent study \cite{apolo2009}.
In a further study, the upper critical dimension $d_u$ of the NWP model was searched for by performing simulations on hyper-cubic lattice graphs in dimensions $d\!=\!2$ through $7$ \cite{Oli10} and evidence was found for an upper critical dimension $d_u\!=\!6$. 
In contrast to that, another upper critical dimension $d_u^{\text{DPL}}\!=\!3$ was determined for densely packed loops which appear if the disorder parameter is far above the critical point that indicates the phase transition \cite{Oli11}.
Only recently, the mean-field behavior of the NWP model was investigated on a random graph with fixed connectivity \cite{Oli11meanField}.
By means of numerical simulations as well as an analytic approach (using the replica symmetric cavity method for a related polymer problem),
the location of the phase transition and the values of the critical exponents were determined.

Here we study the geometrical properties of minimum-weight paths connecting opposite (free) boundaries in the 2D NWP model at the critical point.
To be more precise, we find the answer to the question whether these paths belong to the particular family of planar curves that can be described in terms 
of SLE \cite{cardy2005,kager2004,bauer2006}.
The (chordal) SLE formulation describes a self-avoiding curve $\gamma_t$ 
(``time'' $t\in \mathbb{R}^+$) in the upper half 
plane $\bf{H}$, which is 
equivalent to any other simply connected planar domain by conformal invariance.
The curve $\gamma_t$ grows dynamically from the origin to infinity with time $t$.
A conformal mapping $g_t(z)$ ($z\in \mathbb{C}$ corresponding to the 
two-dimensional plane)
maps the complement of the set consisting of $\gamma_t$ plus all points that are not reachable from infinity without crossing $\gamma_t$, i.e.\ the hull, to the upper half plane at each infinitesimal time step.
Thus, the boundary between reachable and unreachable points
is mapped on the real axis and $g_t(z)$ maps the growing tip 
successively to a real point $a_t$. 
As a consequence, $a_t$ moves continuously on the real axis during the growing process.
Then the time-evolution of $g_t(z)$ satisfies an ordinary differential equation, also referred to as Loewner equation:
\begin{equation}
  \frac{d g_t(z)}{dt}=\frac{2}{g_t(z)-a_t}.
  \label{eq:loewner}
\end{equation}
If $a_t$ is proportional to an one-dimensional Brownian motion $\sqrt{\kappa}B_t$, which is a stochastic Markov process, the resulting curve will be conformal invariant \cite{schramm2000} and the evolution will be referred to as SLE.
Consequently an ensemble of curves that can be described in terms of SLE is characterized and classified by just one parameter, the so-called \emph{diffusion constant} $\kappa$.
In the last decade, a lot of physical systems have been studied in the context of SLE.
To name but a few, boundaries of vorticity clusters in 2D turbulence ($\kappa\!=\!6$) \cite{bernard2006}, spin cluster boundaries in the Ising model at the critical temperature ($\kappa\!=\!3$) \cite{smirnov2006}, and critical percolation ($\kappa\!=\!6$) \cite{smirnov2001} fall into the classification scheme of SLE.
Numerical evidence \cite{amoruso2006} indicates that
 also two-dimensional spin-glasses can be described in terms of SLE.
In contrast to these examples, as it will be shown in this article,
this is not the case for the NWP model.

The remainder of the presented article is organized as follows.
In Section \ref{sect:model}, we introduce the model in more detail and outline the algorithm used to compute the minimum-weight configurations consisting of loops plus one path.
In Section \ref{sect:results}, we describe two measuring techniques for determining the diffusion constant and present the results of our numerical simulations.
In Section \ref{sect:conclusions} we conclude with a summary.
Note that an extensive summary of this paper is available at the \textit{papercore database} \cite{papercore}.


\section{Model and Algorithm\label{sect:model}}

In this article we study 2D square lattice graphs $G\!=\!(V,E)$ with both rectangular and circular shape. 
Nevertheless, in this section the model and the algorithm are described on the basis of the rectangular form only, since there is no conceptual distinction between either shapes.
Some details concerning the different shapes
 are outlined in Section \ref{sect:results}.

The top and bottom
boundaries of the rectangular square lattice graphs are chosen to be free and the left and right
 ones are periodic, i.e.\ the considered lattices are cylindrical.
The system size $L=L_{\rm y}$ denotes the number of nodes in the vertical direction.
As justified in Section \ref{sect:results} the number of nodes in the 
horizontal profile corresponds to $L_{\rm x}=4L\!+\!1$.
Thus the considered graphs have $N\!=\!|V|\!=\!L(4L\!+\!1)$ sites $i\!\in\!V$ and a number of $|E|\!=\!(2L\!-\!1)(4L\!+\!1)$ undirected edges $\{i,j\}\!\in\!E$ that join adjacent sites $i,j\!\in\!V$.
We further assign a weight $\omega_{ij}$ to each edge contained in $E$.
These weights represent quenched random variables that introduce disorder to the lattice and are drawn from a distribution that allows for values of either sign.
In the presented article the edge weights are taken from a ``Gauss-like'' distribution characterized by a tunable disorder parameter $\rho$:
\begin{align}
	P(\omega)=(1-\rho)&\delta(\omega-1)+\rho\exp(-\omega^2)/\sqrt{2\pi},\nonumber\\&\hspace{3cm}0\leq\rho\leq1.
	\label{eq:gauss}
\end{align}
This weight distribution explicitly allows for loops $\mathcal{L}$ with negative weight given by $\omega_{\mathcal{L}}\!\equiv\!\sum_{\{i,j\}\in\mathcal{L}}\omega_{ij}$.
In addition one path $\mathcal{P}$ 
is forced on the lattice whose endpoints lie on the  horizontal boundaries.
One endpoint is fixed at the central node of the lower boundary and the 
other end terminates at a site at the top boundary determined
by the optimality criterion given below.
Moreover, the loops and the path are not allowed to intersect each other, hence every edge is contained in at most one loop (or path).
Given a realization of the disorder, the exact geometrical configuration $\mathcal{C}$ of loops and one path is determined by the following criterion:
The configurational energy $\mathcal{E}$ defined as
\begin{equation}
	\mathcal{E}=\sum_{\mathcal{L}\in\mathcal{C}}\omega_{\mathcal{L}}
+\omega_{\mathcal{P}}
	\label{eq:energy}
\end{equation}
has to be minimized, where $\omega_{\mathcal{P}}
$ denotes the total of all edge weights belonging to the path.
Such minimum-weight configurations for different values of the disorder parameter are presented in Fig.\ \ref{fig:alltogether}.


To get over that ``minimum-weight configuration problem'' we transform 
the original graph as detailed in \cite{ahuja1993} to an appropriate 
auxiliary graph and detect a \emph{minimum-weight perfect matching} 
(MWPM) \cite{cook1999,opt-phys2001,melchertThesis2009} subsequently.
Here, we give a concise description of the algorithmic procedure 
illustrated in Fig.\ \ref{fig:algorithm} for a given realization of the 
disorder (for clarity without periodic boundary conditions):

\begin{figure}[bt]
\centerline{
\includegraphics[width=1.\linewidth]{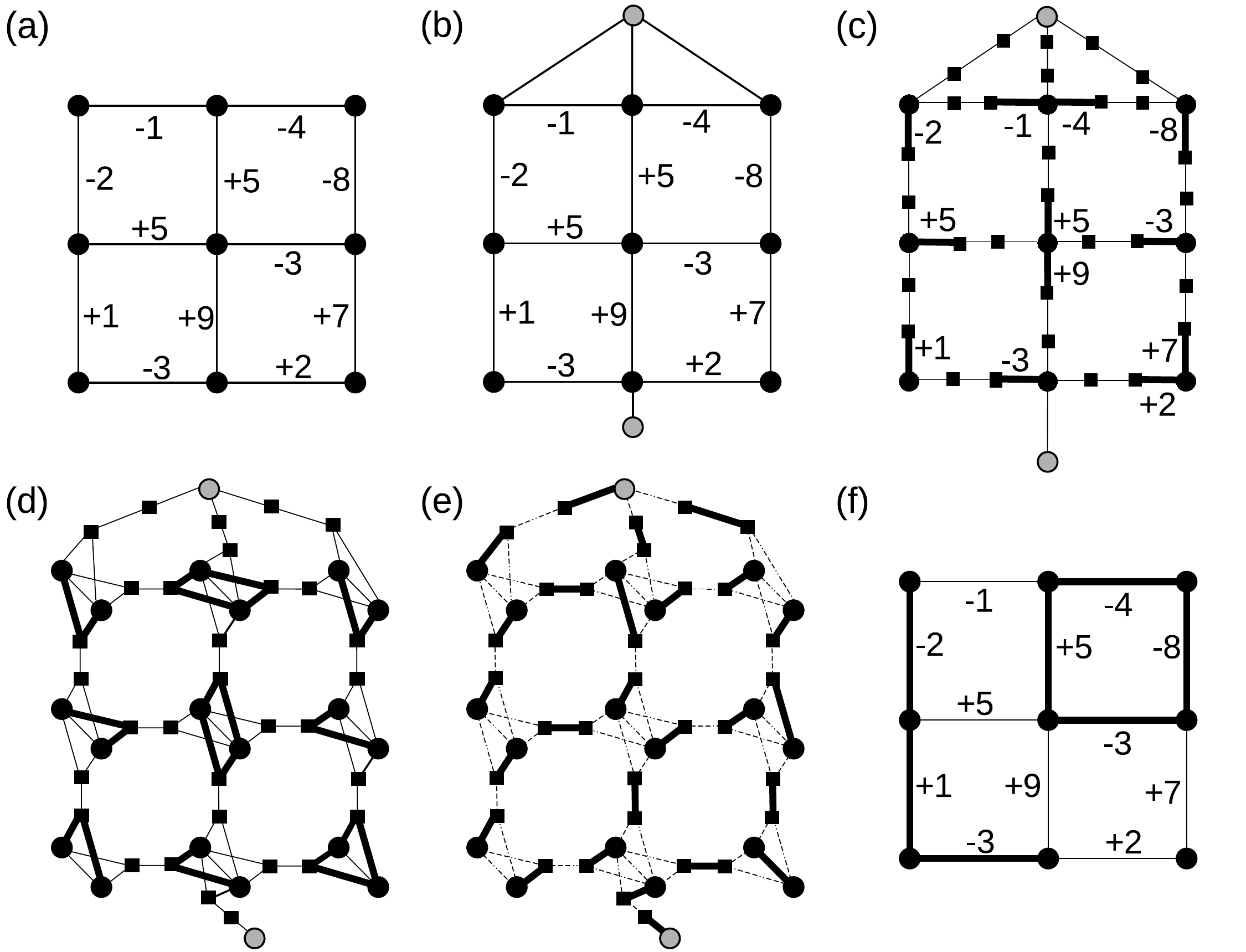}}
\caption{
Illustration of the algorithmic procedure:
Illustrated edges that are not marked by a weight visually, carry a zero weight actually (except in (d) and (e)).
(a) The initial, weighted lattice graph. 
(b) Two extra nodes (gray) are added to render a path possible.
They are connected to the lattice boundaries as described in the text.
(c) Each edge is replaced by two nodes (black boxes) and three edges joining the original nodes and the added nodes in a row.
The initial weights are assigned to one of the added edges (bold edges) that join one of the original nodes and one extra node.
(d) The original nodes are duplicated preserving their neighbors.
Additionally, the duplicated nodes are interconnected by an added edge carrying a zero weight.
(e) A MWPM is computed: bold edges are matched and dashed ones are unmatched.
(f) Reconstruction to the initial lattice taking the MWPM result into account.
The thickly denoted path and loop represent the minimally weighted configuration.
\label{fig:algorithm}}
\end{figure}  

(1) In order to force a path that satisfies the specified boundary conditions, two ``extra'' nodes are added in the following way:
One extra node is linked to the node located in the center of the lower boundary and the other extra node is connected to all nodes at the upper boundary (c.f.\ Fig.\ \ref{fig:algorithm}(b)).
All added edges get zero weight.

(2) Each edge is replaced by two ``additional''
 nodes and three edges which are arranged in a row as depicted in Fig.\ \ref{fig:algorithm}(c).
Therein, one of the two edges connecting an additional node to an original node gets the same weight as the corresponding edge in the original graph $G$.
The remaining two edges get zero weight.

(3) Subsequently, the original nodes $i\!\in\!V$ 
(i.e., not the two extra nodes)
are ``duplicated'', i.e.\ $i\!\rightarrow\!i_1,i_2$, along with all their incident edges and the corresponding weights.
For each of these pairs of duplicated nodes, one edge $\{i_1,i_2\}$ with zero weight is added that connects the two nodes $i_1$ and $i_2$. 
The resulting auxiliary graph $G(V_{\rm A},E_{\rm A})$ is illustrated in Fig.\ \ref{fig:algorithm}(d) without showing the edge weight assignment for reasons of clarity.
Note that while the original graph is symmetric, the transformed graph is not.
This is due to the details of the mapping procedure and the particular weight assignment we have chosen.
A more extensive description of the mapping (in terms of minimum-energy domain wall calculations for the 2D Ising spin glass) can be found in Ref.\ 
\cite{melchert2007}.

(4) A MWPM on the auxiliary graph is determined via exact combinatorial optimization algorithms \cite{comment_cookrohe}. 
A MWPM is a minimum-weighted subset $M$ of $E_{\rm A}$, such that each node contained in $V_{\rm A}$ is met by precisely one edge in $M$.  
This is illustrated in Fig.\ \ref{fig:algorithm}(e), where the solid edges represent $M$ for the given weight assignment. The dashed edges are not matched. 
Due to construction, the auxiliary graph consists of an even number of nodes and due to the fact that pairs of duplicated nodes are connected by additional edges (see step (3)), it is guaranteed that a perfect matching exists.

(5) Finally it is possible to find a relation between the matched edges $M$ on $G_{\rm A}$ and a minimally weighted configuration consisting of loops and one path in $G$ by tracing back the steps of the transformation. 
As regards this, note that each edge contained in $M$ that connects an additional node (square) to a duplicated node (circle) corresponds to an edge in $G$ that is part of a loop or path, see Fig.\ \ref{fig:algorithm}(f).
More precisely, there are always two such edges in $M$ that correspond to one loop/path segment on $G$. 
All the edges in $M$ that connect like nodes (i.e.\ duplicated-duplicated, or additional-additional) carry zero weight and do not contribute to the 
minimally weighted configuration and correspondingly do not belong to
loops or to the path in the original graph.
Once such a configuration is found, a depth-first search \cite{ahuja1993,opt-phys2001} can be used to identify the path and to determine its geometrical properties. 
For the weight assignment illustrated in Fig.\ \ref{fig:algorithm}(a), the path features $\omega_p\!=\!-4$ and length $\ell\!=\!3$.

It is important to emphasize that the result of the calculation is a collection $\mathcal{C}$ of loops and one path, such that their total weight, and consequently the configurational energy $\mathcal{E}$, is minimized. 
Hence, one obtains a global collective optimum. 
This implies that the resulting path is not necessarily the minimally weighted path on the lattice: due to the no-crossing condition, 
loops might affect its precise location. 
While all loops that contribute to $\mathcal{C}$ possess a negative weight, the weight of the path might be positive as well.
The set of loops might be empty, but the addition of the two extra nodes during the first step of the mapping outlined above ensures 
that the solution of the auxiliary minimum-weight perfect matching problem always yields (at least) a path.\\

Note that the choice of the weight assignment in step (2) is not unique, i.e.\ there are different ways to choose a weight assignment that all result in equivalent sets of matched edges in the transformed lattice, corresponding to the minimum-weight configuration in the original lattice.
Some of these weight assignments result in a more symmetric transformed graph (see, e.g.\ \cite{ahuja1993}).
However, this is only a technical issue that does not affect the result.


\section{Results \label{sect:results}}
\begin{figure}[bt]
\centerline{
\includegraphics[width=1.\linewidth]{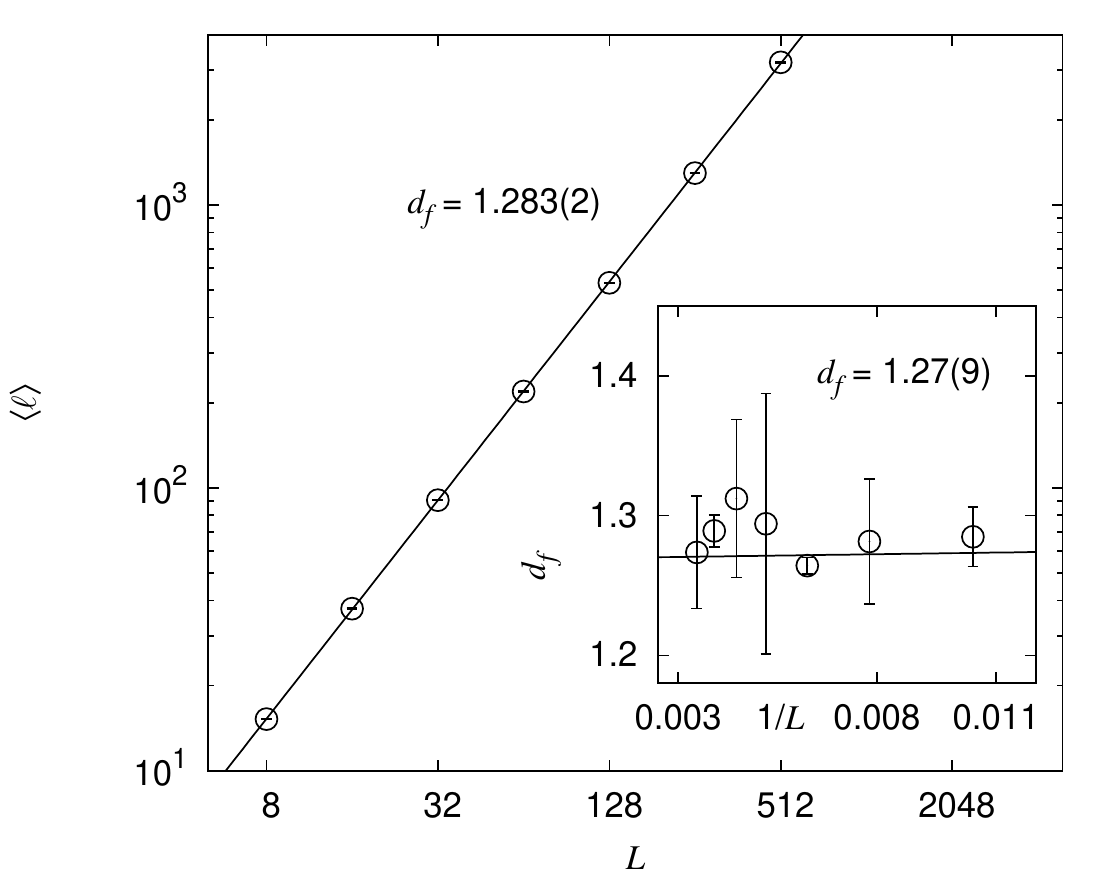}}
\caption{
Estimation of the fractal dimension $d_f$ by means of a simple fit to
 the power law data. 
The inset illustrates the estimation using local effective exponents.
Both estimates go well together within the error bars.
\label{fig:fracDim}}
\end{figure}

As pointed out in the introduction, we aim at finding an answer to the question whether paths in the NWP model can be described in terms of SLE at the critical point, and, if so, which value of $\kappa$ characterizes these paths.
The critical point $\rho_c\!=\!0.3413(9)$ has been estimated in the same way as in Ref.\ \cite{melchert2008} with slightly higher precision.
Principally SLE curves can be calculated by Eq.\ \ref{eq:loewner}.
This stochastic differential equation does not provide a unique curve for a given diffusion constant $\kappa$, but provides a set of curves that exhibits particular properties in a statistical sense.
These properties can be utilized to approach the question whether a given ensemble of curves is a potential candidate for showing SLE.\\
The first property takes the fractal dimension $d_f$ into account. 
The fractal dimension can be defined from the scaling of the average length $\langle\ell\rangle$ of the paths as a function of the system size $L$, according $\langle\ell\rangle\!\sim\!L^{d_f}$.
In the context of SLE there is a relation between the diffusion constant and the fractal dimension \cite{beffara2004, beffara2008}:
\begin{equation}
	\kappa = 8(d_f-1),\hspace{0.7cm}\kappa\leq8.
	\label{eq:fracDim}
\end{equation}
The fractal dimension is fixed at $d_f\!=\!2$ for $\kappa\!>\!8$.
A simple fit to the power-law data shown in Fig.\ \ref{fig:fracDim} yields the exponent $d_f\!=\!1.283(2)$. 

\begin{figure*}[ht]
\centerline{
\includegraphics[width=1\linewidth]{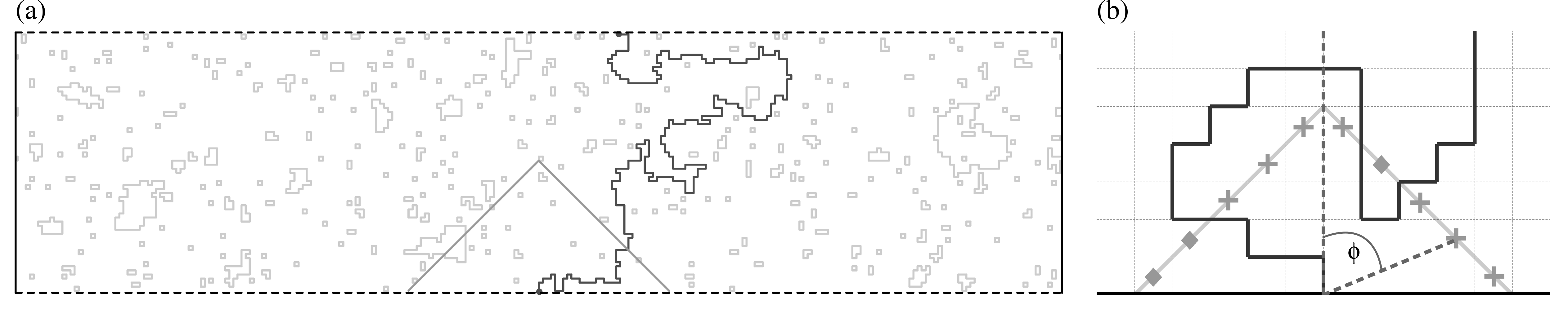}}
\caption{
(a) Illustration of a minimum-weight configuration consisting of loops (gray) and one path (black) in a lattice of size $257\!\times\!64$. 
Due to the ``Manhattan-structure'' of the underlying lattice, the semicircle ($R\!=\!32$) assumes the shape of a triangle (light gray). 
The boundaries in vertical (bottom-top) direction (dashed) 
are free and the  ones 
in horizontal direction (solid) are periodic. 
The edge weights are taken from a ``Gauss-like'' distribution shown in Eq.\ 
(\ref{eq:gauss}) featuring $\rho\!=\!0.3413$. 
The path is forced to connect the horizontal boundaries as described in the text. 
The minimum-weight configuration has to minimize Eq.\ (\ref{eq:energy}).
(b) Sketch of a lattice that approximates the upper half plane to check SLPF predictions at some certain places located on the semicircle.
The checkpoints marked by rhombuses lie on the left and the other ones (crosses) on the right of the path (black).
For every realization of the disorder, the predictions of SLPF are examined independently for each checkpoint.
$\phi$ states the angle between the ordinate and considered checkpoints in the upper half plane.
\label{fig:measure}}
\end{figure*}  

An estimation by extrapolation using local effective exponents \cite{havlin2004} yields $d_f\!=\!1.27(9)$. 
Both estimates go well together within the error bars.
Each data point is generated by means of $12800$ realizations of the disorder.
According to Eq.\ (\ref{eq:fracDim}), 
the more precise estimate yields $\kappa\!=\!2.26(2)$.

\begin{figure}[bt]
\centerline{
\includegraphics[width=1.\linewidth]{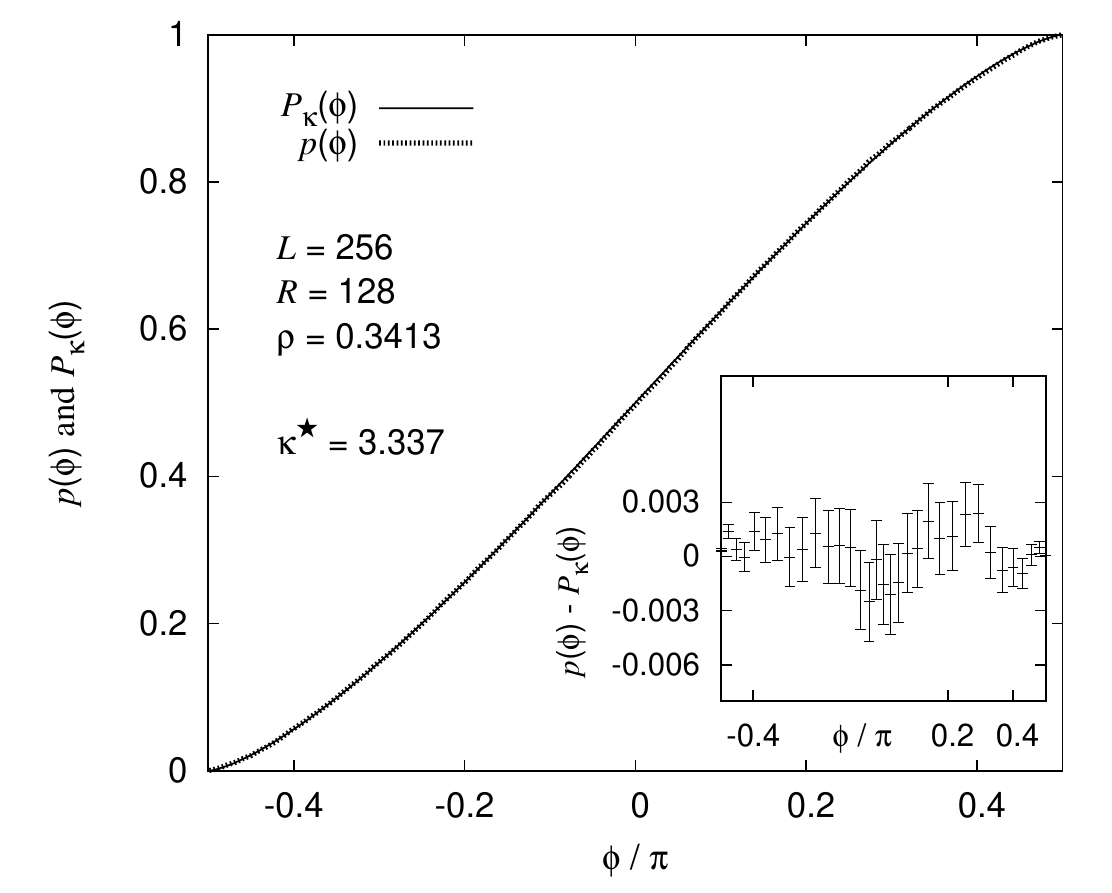}}
\caption{ Comparison between the measured probabilities $p(\phi)$ and
the predictions of SLPF $P_\kappa(\phi)$.  The left passage
probability $p(\phi)$ has been measured for various points on a
semicircle ($R\!=\!128$).  $\phi$ denotes the angle that is
illustrated in Fig.\ \ref{fig:measure} (right).  The computational
simulations have been performed at the critical point in a lattice of
size $1025\!\times\!256$ using $51200$ realizations of the disorder.
The chosen diffusion constant $\kappa^\star\!=\!3.337$ provides the finest
agreement between $P_\kappa(\phi)$ and $p(\phi)$ (cf.\ Fig.\
\ref{fig:squaredDev}).  The inset illustrates the deviation of the
measured left-passage probability from the SLE predictions.  For
reason of clarity, merely a few calculated data points are
illustrated.  As a matter of fact, $p(\phi)$ has been estimated for
$257$ different values of $\phi$.
\label{fig:comp}}
\end{figure}

Similar to \cite{schwarz2009,roma2008,picco2008,chatelain2010,bernard2007} we examine predictions of Schramm's left passage formula (SLPF) in order to obtain a second estimation for $\kappa$.
As proven in \cite{schramm2001}, 
\begin{align}
	P_\kappa(z) = &\frac{1}{2}+\frac{\Gamma(4/\kappa)}{\sqrt{\pi}\Gamma((8-\kappa)/2\kappa)}\nonumber\\&\hspace{0.2cm}\times{}_2F_1\left(\frac{1}{2},\frac{4}{\kappa};\frac{3}{2};-\left(\frac{x}{y}\right)^2\right)\frac{x}{y}	
	\label{eq:slpf}
\end{align}
yields the probability that a curve which can be described in terms of SLE will pass to the left of a given point $z\!=\!x\!+\!iy$ in the upper half plane provided that the curve links the origin to infinity.
${}_2F_1$ denotes the Gaussian hypergeometric function and $\Gamma$ the gamma function.
$P_\kappa(z)$ does not depend on the distance between the origin and point $z$.
Thus the ratio between $\text{Re}(z)$ and $\text{Im}(z)$ can be replaced by a function of an angle.
A proper angle $\phi\!\in\![-\pi/2,\pi/2]$ between the imaginary axis and the reference point $z$ is located in the origin of the upper half plane as depicted in Fig.\ \ref{fig:measure} (b) \cite{schwarz2009}.
Thus the ratio between $x$ and $y$ is given by $x/y\!=\!\tan\phi$.
A curve $\gamma$ will be considered as passing to the left of a given point $z$ if there exists a continuous path between $z$ and the positive real axis that does not intersect $\gamma$ \cite{cardy2007}.
To support intuition, Fig.\ \ref{fig:comp} shows SLPF for a particular value of $\kappa$ (at this point the reference to the figure 
is only meant to illustrate the general functional form of the SLPF, the precise value of the parameters $\rho$ and $\kappa$ and
the results implied by them are discussed below).

Note that Eq.\ (\ref{eq:slpf}) holds in the upper half plane only. 
Subsequently we consider two different lattice geometries that aim to 
approximate the upper half plane.

\begin{figure*}[tb]
\centerline{
\includegraphics[width=0.9\linewidth]{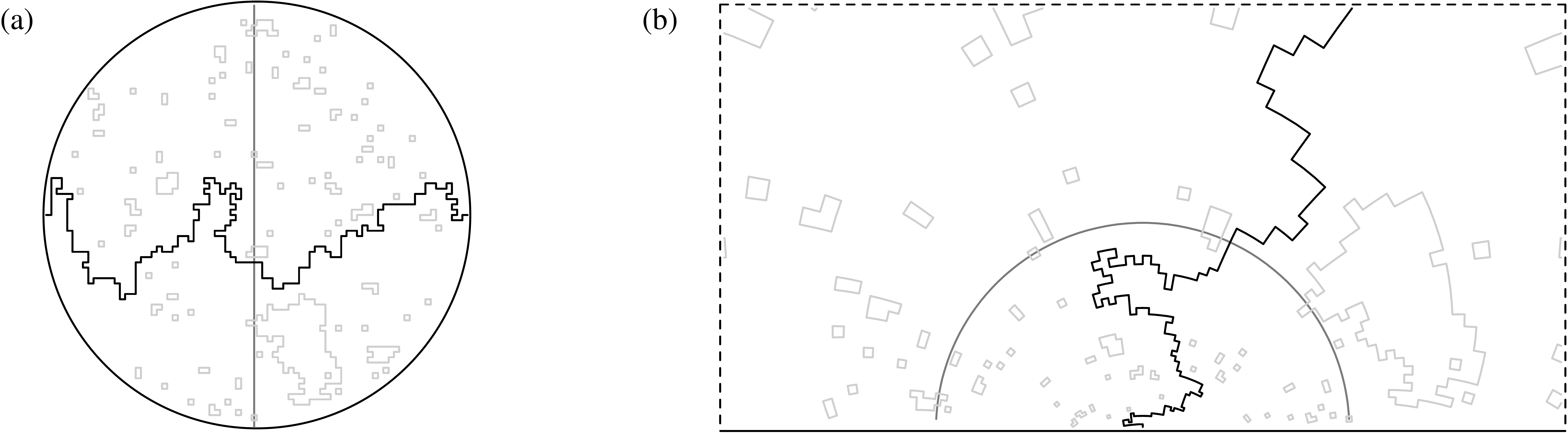}}
\caption{ (a) Illustration of a minimum-weight configuration
consisting of loops (gray) and one path (black) in the unit disc with
a radius of $40$.  The terminal points of the path are fixed on
$-1+0i$ and $1+0i$ to ensure that they will be mapped as required by
Cayley's function.  All points which Schramm's left passage formula is
checked for are located on the vertical solid line (deep
gray).\newline (b) The same minimum-weight configuration as in (a)
after the conformal mapping by Cayley's function has been performed,
so the path connects the origin to infinity in the upper half plane.
The dashed lines indicate that the lattice is not displayed
completely.  Due to the mapping, the original, vertical line
containing all checkpoints becomes a semicircle.
\label{fig:discHp}}
\end{figure*}

On the one hand we approximate the upper half plane directly by constructing a rectangular square lattice featuring appropriate boundary conditions.
On the other hand we design a square lattice exhibiting a circular contour and regard it as unit disc $\bf{E}$ in the complex plane.
The unit circle can be mapped conformally into the upper half plane by a simple mapping called Cayley function \cite{cayley1846}:
\begin{equation}
	g:\bf{E} \rightarrow \bf{H}, \hspace{0.6cm} z \mapsto i\,\frac{1+z}{1-z}.
\label{eq:cayley}
\end{equation}
The mapping provides $g(-1)\!=\!0$ and $g(1)\!=\!\infty$, thus a curve in $\bf{E}$ linking $-1\!+\!0i$ and $1\!+\!0i$ is mapped by $g$ as required.
For a given realization of the disorder, the ground state in the unit disc and the corresponding ground state in the upper half plane are illustrated in Fig.\ \ref{fig:discHp} to support intuition.

\begin{figure}[bt]
\centerline{
\includegraphics[width=1.\linewidth]{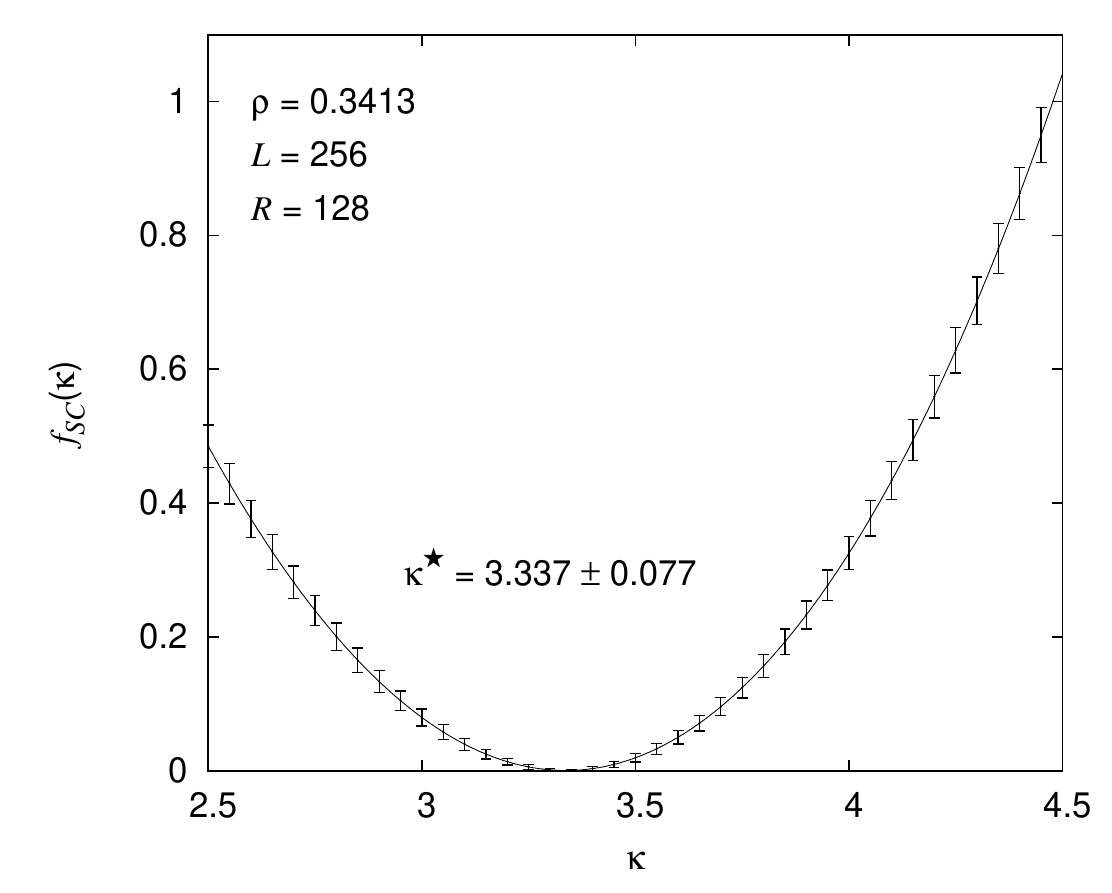}}
\caption{
The cumulative squared deviation $f_{SC}(\kappa)$ of the measured probabilities $p(\phi)$ from the corresponding values given by Schramm's left passage formula $P_\kappa(\phi)$ as a function of the diffusion constant $\kappa$.
The semicircle, which contains the checkpoints, features a radius $R\!=\!128$ in a lattice of size $1025\!\times\!256$.
The simulations ($51200$ realizations of the disorder) have been performed at the critical point.
The squared deviation has been calculated for discrete values of $\kappa$ between $2.5$ and $4.5$ with step range $0.001$.
For reasons of clarity, only a few error bars are displayed.
The cumulative squared deviation yields the finest agreement between $p(\phi)$ and $P_\kappa(\phi)$ at $\kappa^\star\!=\!3.337\pm0.077$.
\label{fig:squaredDev}}
\end{figure}

First we consider the rectangular lattice shape.
In order to approximate the upper half plane as accurately as possible, the terminal points of the paths in the NWP model are chosen as described in Section \ref{sect:model}.
The terminal point located at the central node of the lower boundary is considered as the origin of the coordinate system.
Furthermore, all considered lattices are on a scale of $4\!:\!1$ to avoid the influence of the finite size in horizontal direction.
An example of a minimally weighted configuration in such a lattice is shown in Fig.\ \ref{fig:measure} (a) for a lattice of size $257\!\times\!64$.
Although SLPF applies for all points in the upper half plane, its predictions are often checked for properly chosen points only.
Aside from $\kappa$, Eq.\ (\ref{eq:slpf})
 depends on angle $\phi$ merely, thus it is sufficient to consider points located on a semicircle, whose central point is located in the origin of the upper half plane.
The distance between the center of this semicircle and every point located on its boundary should be equal.
For that reason and due to the ``Manhattan structure'' of the underlying lattice, the semicircle assumes the shape of a triangle as illustrated in Fig.\ \ref{fig:measure}.
The path is able to cross the semicircle at particular points only.
The checkpoints we consider to examine predictions of SLPF are located exactly between these possible crossing points.
Each checkpoint is examined independently for a given realization of the disorder, i.e.\ for each given path.
In so doing, for a sufficient large number of samples $N_s$, each checkpoint will exhibit a measured left-passage probability $p(\phi)\pm\sqrt{p(\phi)(1\!-\!p(\phi))/(N_s\!-\!1)}$ \cite{practicalGuide2009}.
A comparison between these measured probabilities $p(\phi)$ and the corresponding predictions of SLPF provides an estimate for $\kappa$.
The minimum of the cumulative squared deviation 
\begin{equation}
	f_{SC}(\kappa) = \sum_{\phi\in SC}\left[p(\phi)-P_\kappa(\phi)\right]^2
	\nonumber
\end{equation}
yields the finest agreement between $P_\kappa(\phi)$ and $p(\phi)$, where $SC$ denotes a set containing all possible angles of the considered checkpoints.
The cumulative squared deviation as a function of $\kappa$ is illustrated in Fig.\ \ref{fig:squaredDev} for a lattice of size $1025\!\times\!256$.

\begin{figure}[tb]
\centering
\includegraphics[width=1.0\linewidth]{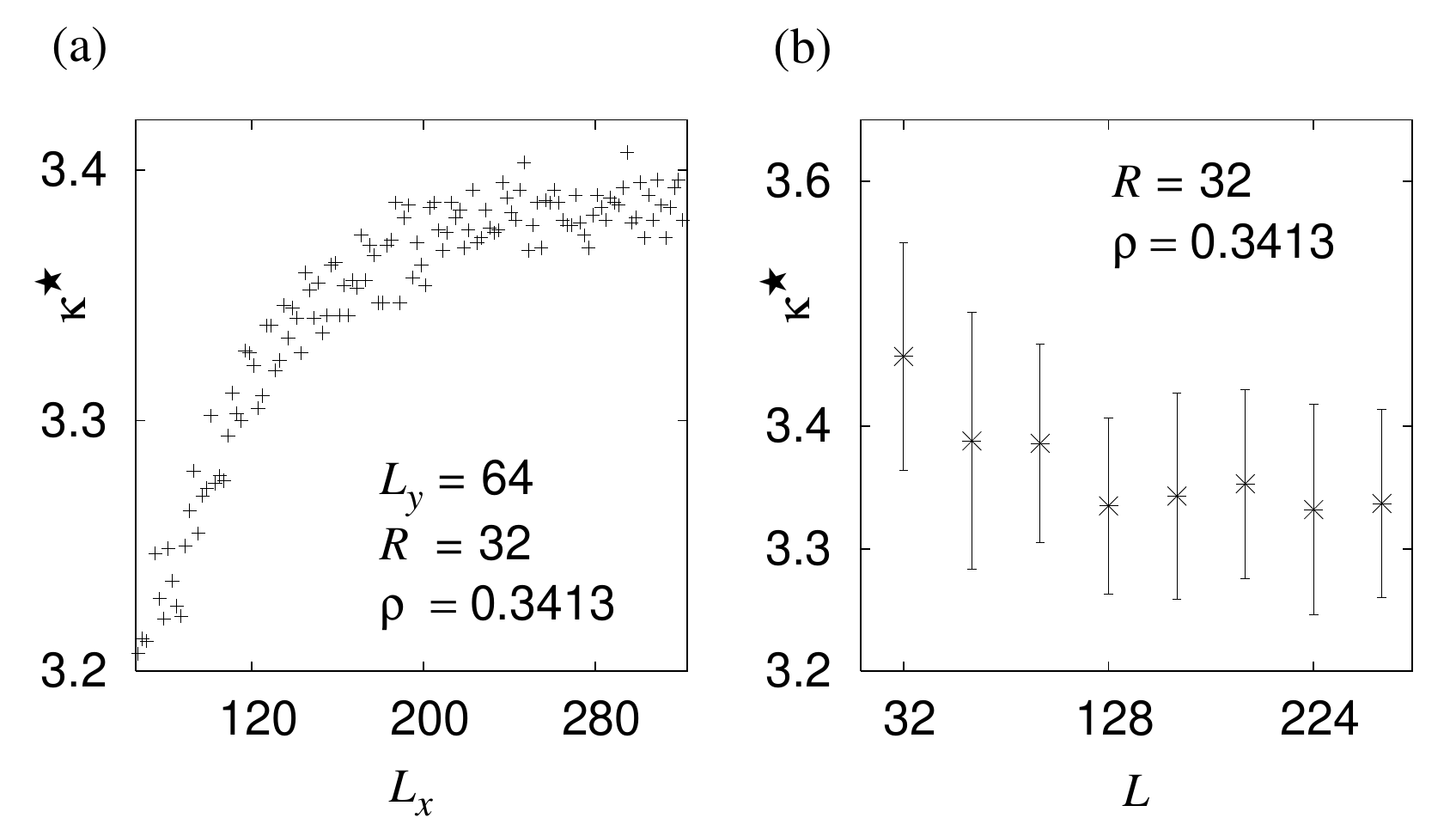}
\caption{
(a) The diffusion constant $\kappa^\star$ depending on the ratio between the lateral extensions of the lattice.
The horizontal extension $L_{\rm x}$ is varied in length and the radius $R$ of the semicircle is chosen to be the half of the system size $L=L_{\rm y}$.
The estimates of the diffusion constant have been determined at the critical point by means of $51200$ realizations of the disorder.
For reasons of clarity, the error bars are not illustrated.
Note that all estimates (except for the shortest horizontal extensions until 80) are compatible to each other within the error bars.\newline
(b) The diffusion constant $\kappa^\star$ depending on the system size $L$ that denotes the number of nodes in the vertical direction of the lattice.
The horizontal extension is equivalent to $4L\!+\!1$.
The radius of the semicircle is chosen to be the half of the system size in every cases.
The estimates of the diffusion constant have been determined at the critical point by means of $51200$ realizations of the disorder.
\label{fig:upperHP}}
\end{figure}  


In so doing $f_{SC}(\kappa)$ and its error bar $\Delta f_{SC}(\kappa)$ have been calculated for a set of $M=4000$ individual values of $\kappa\in K$, where $K=\{\kappa_1,\kappa_2,\ldots,\kappa_M\}$.
If $\kappa^\star$ denotes the diffusion constant where the squared deviation 
$f_{SC}(\kappa)$ is minimal, its error bar $\Delta \kappa^\star$ will be estimated as follows: 
Defining $\Omega\!=\!\{ \kappa\!\in\!K \,\mid\, f_{SC}(\kappa^\star)\!
+\!\Delta f_{SC}(\kappa^\star)\ge f_{SC}(\kappa)\!-\!\Delta f_{SC}(\kappa) \}$,
 one has $\Delta \kappa^\star\!=\!\max \{ | \kappa^\star\!-\!\kappa | \,\mid\,\kappa\!\in\!\Omega\}$. This results in 
$\kappa^\star\!=\!3.337\pm0.077$. One might wonder
whether the result is affected by different lattice-size effects, which we
consider now.

As mentioned above, we design the underlying lattice on scale of $4\!:\!1$ to approximate the upper half plane. 
It is evident from Fig.\ \ref{fig:upperHP}(a) that this ratio is sufficient.
We estimate the diffusion constant using the squared deviation technique described above for different aspect ratios.
In doing so we fixed the number $L_y=L$ of nodes in the vertical direction 
and varied the horizontal extension $L_x$. 
It becomes apparent that the diffusion constant does not vary significantly above a horizontal extension that is four times bigger than the vertical extension of the lattice, in particular, all values
are clearly different from the value corresponding to the fractal dimension.

In order to receive an impression of the influence of finite size effects, 
we varied the system size $L$ (using the 4:1 aspect ratio) 
as depicted in Fig.\ \ref{fig:upperHP}(b). 
Bear in mind that the system size $L$ corresponds to the number of nodes in the vertical lattice direction.
It becomes apparent that the diffusion constant estimated by using the squared deviation technique is steady sufficiently atop $L\!=\!128$.
A further study, which is not illustrated here, reveals that $\kappa^\star$ does not vary significantly for differently chosen radii of the semicircle.

Thus a good approximation to the upper half plane is provided by a lattice of size $1025\!\times\!256$ in addition to a semicircle with radius $R\!=\!128$.
So we estimated $\kappa$ for such a lattice by examining SLPF predictions. 
As mentioned above,
the cumulative squared deviation $f_{SC}(\kappa)$ of $p(\phi)$ and $P_\kappa(\phi)$ at $\rho_c$ yields $\kappa^\star\!=\!3.337(77)$ (cf.\ Fig.\ \ref{fig:squaredDev}).
A comparison to the estimate obtained from Eq.\ (\ref{eq:fracDim})
 ($\kappa\!=\!2.26(2)$) reveals clearly that both estimates do not agree.

To ensure that this conspicuous distinction does not arise from the 
approximation of the upper half plane by designing a rectangular 
lattice, we also examine paths on the unit disc and check the predictions 
of SLPF after mapping the unit disc (and, consequently, the determined 
ground states) into the upper half plane, as depicted
in Fig.\ \ref{fig:discHp}.
The radius of the disc is discretized by $128$ nodes and the terminal 
points of the paths are set as described above (cf.\ \ref{fig:discHp}(a)).
Just like before, the predictions of SLPF are examined merely for 
checkpoints located on a semicircle (cf.\ Fig.\ \ref{fig:discHp}(b)) 
in an analogous manner.
Taking $12800$ realizations of the disorder into account, 
the diffusion constant $\kappa^\star\!=\!3.34(14)$ provides the best 
match to the measured left-passage probability.
Since this estimation of $\kappa$ is equal to the estimation 
obtained before, it becomes apparent that the upper half 
plane has been approximated properly by the rectangular lattice and 
the estimate of $\kappa$ obtained by Eq.\ 
(\ref{eq:fracDim}) differs significantly from the estimate 
considering SLPF as an actual fact.
Consequently this means that paths in the NWP model cannot be 
described in terms of SLE.

Note that we also compared
 the measured left-passage probability $p(\phi)$ and 
the corresponding predictions of SLPF directly.
From Fig.\ \ref{fig:comp} it is evident that $p(\phi)$ and 
$P_{3.337}(\phi)$ go well together at almost all value of $\phi$.
Although the paths of the NWP model do not show SLE properties, 
the predictions of $P_{3.337}(\phi)$ appear to apply for this model 
at the critical point.


\section{Conclusions \label{sect:conclusions}}
We examined whether paths in the negative-weight percolation model fall into the classification scheme of SLE.
In order to address this question, we studied their geometrical properties and compare them to predictions that arises from the SLE theory.
First we determine the fractal dimension $d_f\!=\!1.282(2)$ by measuring the average path length as a function of the system size.
The relation $\kappa\!=\!8(d_f\!-\!1)$ provides an estimate for the diffusion constant $\kappa\!=\!2.26(2)$.
In order to get a second estimate for $\kappa$, we examined Schramm's left passage formula, which states the probability that a curve featuring particular boundary conditions and showing SLE passes to the left of a given point in the upper half plane.
For this purpose we designed lattices featuring different shapes in order to approximate the upper half plane as sufficiently as possible.
A comparison between the predictions of Schramm's formula and our numerical results provides the estimate $\kappa^\star\!=\!3.34(8)$.
Because of the conspicuous distinction between both estimates of $\kappa$, it became apparent that paths in the NWP model cannot be described in terms of SLE.
We propose two possible reasons for this failure of SLE
for the NWP. First, the minimum-weight path from the center bottom
to the top might be affected by negative loops, which arise due
to the minimality criterion. The edges occupied by these
loops are not available to the minimum-weight path. Nevertheless, as visible
from Fig.\ \ref{fig:alltogether}, these additional loops
are small and somehow rare, hence, the effect might be negligible. 
Note that algorithmically excluding the loops from the minimization
procedure will make the problem algorithmically hard, i.e., the running
time increases exponentially with the system size, such that only very small
instances can be solved.

Second,  the way the minimum-weight path is selected
refers to a global minimization criterion, hence it cannot grow locally.
Loosely speaking, the path can afford to cross a region containing
few negative edges if it arrives to a region which exhibits a high
concentration of negative edges as compensation. This might violate the
so-called ``Markov-property'' of SLE processes \cite{cardy2005}. On the
other hand, this argument applies only to single-realizations
and might be not relevant for the stochastic ensemble, which is what the
SLE property describes. In fact,
domain-walls in $T=0$ 2d spin glasses are also
obtained by a global optimization procedure, and they seem to obey SLE.
Thus, the reason for the failure of the SLE property for NWP is still not 
totally clear to us.


\begin{acknowledgments}
CN acknowledges financial support from the 
DFG (\emph{Deutsche Forschungsgemeinschaft}) under grant AS 136/2-1.
OM acknowledges financial support from the 
DFG  under grant HA 3169/3-1.
The simulations were performed at the HERO cluster of the University
of Oldenburg, funded by the DFG (INST 184/108-1 FUGG) and the 
ministry of Science and Culture (MWK) of the Lower Saxony State,
and the GOLEM I Cluster for 
Scientific Computing both located at the University of Oldenburg (Germany).
\end{acknowledgments}

\bibliography{research.bib}

\begin{thebibliography}{51}
\expandafter\ifx\csname natexlab\endcsname\relax\def\natexlab#1{#1}\fi
\expandafter\ifx\csname bibnamefont\endcsname\relax
  \def\bibnamefont#1{#1}\fi
\expandafter\ifx\csname bibfnamefont\endcsname\relax
  \def\bibfnamefont#1{#1}\fi
\expandafter\ifx\csname citenamefont\endcsname\relax
  \def\citenamefont#1{#1}\fi
\expandafter\ifx\csname url\endcsname\relax
  \def\url#1{\texttt{#1}}\fi
\expandafter\ifx\csname urlprefix\endcsname\relax\def\urlprefix{URL }\fi
\providecommand{\bibinfo}[2]{#2}
\providecommand{\eprint}[2][]{\url{#2}}

\bibitem[{\citenamefont{Stauffer and Aharony}(1994)}]{stauffer1994}
\bibinfo{author}{\bibfnamefont{D.}~\bibnamefont{Stauffer}} \bibnamefont{and}
  \bibinfo{author}{\bibfnamefont{A.}~\bibnamefont{Aharony}},
  \emph{\bibinfo{title}{{Introduction to Percolation Theory}}}
  (\bibinfo{publisher}{Taylor and Francis, London}, \bibinfo{year}{1994}).

\bibitem[{\citenamefont{Janke and Schakel}({2005})}]{janke2005}
\bibinfo{author}{\bibfnamefont{W.}~\bibnamefont{Janke}} \bibnamefont{and}
  \bibinfo{author}{\bibfnamefont{A.}~\bibnamefont{Schakel}},
  \bibinfo{journal}{{Phys. Rev. E}} \textbf{\bibinfo{volume}{{71}}}
  (\bibinfo{year}{{2005}}).

\bibitem[{\citenamefont{Kardar and Zhang}(1987)}]{kardar1987}
\bibinfo{author}{\bibfnamefont{M.}~\bibnamefont{Kardar}} \bibnamefont{and}
  \bibinfo{author}{\bibfnamefont{Y.~C.} \bibnamefont{Zhang}},
  \bibinfo{journal}{Phys. Rev. Lett.} \textbf{\bibinfo{volume}{58}},
  \bibinfo{pages}{2087} (\bibinfo{year}{1987}).

\bibitem[{\citenamefont{Derrida}(1990)}]{derrida1990}
\bibinfo{author}{\bibfnamefont{B.}~\bibnamefont{Derrida}},
  \bibinfo{journal}{Physica A} \textbf{\bibinfo{volume}{163}},
  \bibinfo{pages}{71} (\bibinfo{year}{1990}).

\bibitem[{\citenamefont{Grassberger}(1993)}]{grassberger1993}
\bibinfo{author}{\bibfnamefont{P.}~\bibnamefont{Grassberger}},
  \bibinfo{journal}{J. Phys. A} \textbf{\bibinfo{volume}{26}},
  \bibinfo{pages}{1023} (\bibinfo{year}{1993}).

\bibitem[{\citenamefont{Buldyrev et~al.}(2006)\citenamefont{Buldyrev, Havlin,
  and E.}}]{buldyrev2006}
\bibinfo{author}{\bibfnamefont{S.~V.} \bibnamefont{Buldyrev}},
  \bibinfo{author}{\bibfnamefont{S.}~\bibnamefont{Havlin}}, \bibnamefont{and}
  \bibinfo{author}{\bibfnamefont{S.~H.} \bibnamefont{E.}},
  \bibinfo{journal}{Phys. Rev. E} \textbf{\bibinfo{volume}{73}},
  \bibinfo{pages}{036128} (\bibinfo{year}{2006}).

\bibitem[{\citenamefont{Pfeiffer and Rieger}(2002)}]{pfeiffer2002}
\bibinfo{author}{\bibfnamefont{F.~O.} \bibnamefont{Pfeiffer}} \bibnamefont{and}
  \bibinfo{author}{\bibfnamefont{H.}~\bibnamefont{Rieger}},
  \bibinfo{journal}{J. Phys.: Condens. Matter} \textbf{\bibinfo{volume}{14}},
  \bibinfo{pages}{2361} (\bibinfo{year}{2002}).

\bibitem[{\citenamefont{Pfeiffer and Rieger}(2003)}]{pfeiffer2003}
\bibinfo{author}{\bibfnamefont{F.~O.} \bibnamefont{Pfeiffer}} \bibnamefont{and}
  \bibinfo{author}{\bibfnamefont{H.}~\bibnamefont{Rieger}},
  \bibinfo{journal}{Phys. Rev. {\bf E}} \textbf{\bibinfo{volume}{67}},
  \bibinfo{pages}{056113} (\bibinfo{year}{2003}).

\bibitem[{\citenamefont{Vachaspati and Vilenkin}(1984)}]{vachaspati1984}
\bibinfo{author}{\bibfnamefont{T.}~\bibnamefont{Vachaspati}} \bibnamefont{and}
  \bibinfo{author}{\bibfnamefont{A.}~\bibnamefont{Vilenkin}},
  \bibinfo{journal}{Phys. Rev. D} \textbf{\bibinfo{volume}{30}},
  \bibinfo{pages}{2036} (\bibinfo{year}{1984}).

\bibitem[{\citenamefont{Scherrer and Frieman}(1986)}]{scherrer1986}
\bibinfo{author}{\bibfnamefont{R.~J.} \bibnamefont{Scherrer}} \bibnamefont{and}
  \bibinfo{author}{\bibfnamefont{J.~A.} \bibnamefont{Frieman}},
  \bibinfo{journal}{Phys. Rev. D} \textbf{\bibinfo{volume}{33}},
  \bibinfo{pages}{3556} (\bibinfo{year}{1986}).

\bibitem[{\citenamefont{Hindmarsch and Strobl}(1995)}]{hindmarsch1995}
\bibinfo{author}{\bibfnamefont{H.}~\bibnamefont{Hindmarsch}} \bibnamefont{and}
  \bibinfo{author}{\bibfnamefont{K.}~\bibnamefont{Strobl}},
  \bibinfo{journal}{Nucl. Phys. {\bf B}} \textbf{\bibinfo{volume}{437}},
  \bibinfo{pages}{471} (\bibinfo{year}{1995}).

\bibitem[{\citenamefont{Cieplak et~al.}(1994)\citenamefont{Cieplak, Maritan,
  and Banavar}}]{cieplak1994}
\bibinfo{author}{\bibfnamefont{M.}~\bibnamefont{Cieplak}},
  \bibinfo{author}{\bibfnamefont{A.}~\bibnamefont{Maritan}}, \bibnamefont{and}
  \bibinfo{author}{\bibfnamefont{J.~R.} \bibnamefont{Banavar}},
  \bibinfo{journal}{Phys. Rev. Lett.} \textbf{\bibinfo{volume}{72}},
  \bibinfo{pages}{2320} (\bibinfo{year}{1994}).

\bibitem[{\citenamefont{Melchert and Hartmann}(2007)}]{melchert2007}
\bibinfo{author}{\bibfnamefont{O.}~\bibnamefont{Melchert}} \bibnamefont{and}
  \bibinfo{author}{\bibfnamefont{A.~K.} \bibnamefont{Hartmann}},
  \bibinfo{journal}{Phys. Rev. B} \textbf{\bibinfo{volume}{76}},
  \bibinfo{pages}{174411} (\bibinfo{year}{2007}).

\bibitem[{\citenamefont{Schwarz et~al.}(2009)\citenamefont{Schwarz,
  Karrenbauer, Schehr, and Rieger}}]{schwarz2009}
\bibinfo{author}{\bibfnamefont{K.}~\bibnamefont{Schwarz}},
  \bibinfo{author}{\bibfnamefont{A.}~\bibnamefont{Karrenbauer}},
  \bibinfo{author}{\bibfnamefont{G.}~\bibnamefont{Schehr}}, \bibnamefont{and}
  \bibinfo{author}{\bibfnamefont{H.}~\bibnamefont{Rieger}},
  \bibinfo{journal}{J. Stat. Mech.} \textbf{\bibinfo{volume}{2009}},
  \bibinfo{pages}{P08022} (\bibinfo{year}{2009}).

\bibitem[{\citenamefont{Stauffer}(1979)}]{stauffer1979}
\bibinfo{author}{\bibfnamefont{D.}~\bibnamefont{Stauffer}},
  \bibinfo{journal}{Phys. Rep.} \textbf{\bibinfo{volume}{54}},
  \bibinfo{pages}{1} (\bibinfo{year}{1979}).

\bibitem[{\citenamefont{Allega et~al.}(1990)\citenamefont{Allega, Fern\'andez,
  and A.}}]{allega1990}
\bibinfo{author}{\bibfnamefont{A.}~\bibnamefont{Allega}},
  \bibinfo{author}{\bibfnamefont{L.~A.} \bibnamefont{Fern\'andez}},
  \bibnamefont{and} \bibinfo{author}{\bibfnamefont{T.}~\bibnamefont{A.}},
  \bibinfo{journal}{Nucl. Phys. B} \textbf{\bibinfo{volume}{332}},
  \bibinfo{pages}{760} (\bibinfo{year}{1990}).

\bibitem[{\citenamefont{Austin et~al.}(1994)\citenamefont{Austin, J., and
  J.}}]{austin1994}
\bibinfo{author}{\bibfnamefont{D.}~\bibnamefont{Austin}},
  \bibinfo{author}{\bibfnamefont{C.~E.} \bibnamefont{J.}}, \bibnamefont{and}
  \bibinfo{author}{\bibfnamefont{R.~R.} \bibnamefont{J.}},
  \bibinfo{journal}{Phys. Rev. D} \textbf{\bibinfo{volume}{49}},
  \bibinfo{pages}{4089} (\bibinfo{year}{1994}).

\bibitem[{\citenamefont{{Cardy}}(2005)}]{cardy2005}
\bibinfo{author}{\bibfnamefont{J.}~\bibnamefont{{Cardy}}},
  \bibinfo{journal}{Ann. of Phys.} \textbf{\bibinfo{volume}{318}},
  \bibinfo{pages}{81} (\bibinfo{year}{2005}), \bibinfo{note}{remark: summary is
  available at \url{http://www.papercore.org/Cardy2005}}.

\bibitem[{\citenamefont{Kager and Nienhuis}(2004)}]{kager2004}
\bibinfo{author}{\bibfnamefont{W.}~\bibnamefont{Kager}} \bibnamefont{and}
  \bibinfo{author}{\bibfnamefont{B.}~\bibnamefont{Nienhuis}},
  \bibinfo{journal}{J. Stat. Phys.} \textbf{\bibinfo{volume}{115}},
  \bibinfo{pages}{1149} (\bibinfo{year}{2004}).

\bibitem[{\citenamefont{Bauer and Bernard}(2003)}]{bauer2006}
\bibinfo{author}{\bibfnamefont{M.}~\bibnamefont{Bauer}} \bibnamefont{and}
  \bibinfo{author}{\bibfnamefont{D.}~\bibnamefont{Bernard}},
  \bibinfo{journal}{Comm. Math. Phys.} \textbf{\bibinfo{volume}{239}},
  \bibinfo{pages}{493} (\bibinfo{year}{2003}).

\bibitem[{\citenamefont{Newman and Barkema}(1999)}]{newman1999}
\bibinfo{author}{\bibfnamefont{M.~E.~J.} \bibnamefont{Newman}}
  \bibnamefont{and} \bibinfo{author}{\bibfnamefont{G.~T.}
  \bibnamefont{Barkema}}, \emph{\bibinfo{title}{Monte {C}arlo Methods in
  Statistical Physics}} (\bibinfo{publisher}{Clarendon Press},
  \bibinfo{address}{Oxford}, \bibinfo{year}{1999}).

\bibitem[{\citenamefont{Schwartz et~al.}(1998)\citenamefont{Schwartz, Nazaryev,
  and Havlin}}]{schwartz1998}
\bibinfo{author}{\bibfnamefont{N.}~\bibnamefont{Schwartz}},
  \bibinfo{author}{\bibfnamefont{A.~L.} \bibnamefont{Nazaryev}},
  \bibnamefont{and} \bibinfo{author}{\bibfnamefont{S.}~\bibnamefont{Havlin}},
  \bibinfo{journal}{Phys. Rev. E} \textbf{\bibinfo{volume}{58}},
  \bibinfo{pages}{7642} (\bibinfo{year}{1998}).

\bibitem[{\citenamefont{Rieger}(2003)}]{rieger2003}
\bibinfo{author}{\bibfnamefont{H.}~\bibnamefont{Rieger}}, \bibinfo{journal}{J.
  Phys. A} \textbf{\bibinfo{volume}{36}}, \bibinfo{pages}{11095}
  (\bibinfo{year}{2003}).

\bibitem[{\citenamefont{Melchert and Hartmann}(2008)}]{melchert2008}
\bibinfo{author}{\bibfnamefont{O.}~\bibnamefont{Melchert}} \bibnamefont{and}
  \bibinfo{author}{\bibfnamefont{A.~K.} \bibnamefont{Hartmann}},
  \bibinfo{journal}{New. J. Phys.} \textbf{\bibinfo{volume}{10}},
  \bibinfo{pages}{043039} (\bibinfo{year}{2008}).

\bibitem[{\citenamefont{Melchert}(2009)}]{melchertThesis2009}
\bibinfo{author}{\bibfnamefont{O.}~\bibnamefont{Melchert}},
  \emph{\bibinfo{title}{{PhD thesis}}} (\bibinfo{publisher}{not published},
  \bibinfo{year}{2009}).

\bibitem[{\citenamefont{Melchert and Hartmann}(2011{\natexlab{a}})}]{oli2011}
\bibinfo{author}{\bibfnamefont{O.}~\bibnamefont{Melchert}} \bibnamefont{and}
  \bibinfo{author}{\bibfnamefont{A.~K.} \bibnamefont{Hartmann}},
  \bibinfo{journal}{Comput. Phys. Commun.} \textbf{\bibinfo{volume}{182}},
  \bibinfo{pages}{1828} (\bibinfo{year}{2011}{\natexlab{a}}).

\bibitem[{\citenamefont{Apolo et~al.}(2009)\citenamefont{Apolo, Melchert, and
  Hartmann}}]{apolo2009}
\bibinfo{author}{\bibfnamefont{L.}~\bibnamefont{Apolo}},
  \bibinfo{author}{\bibfnamefont{O.}~\bibnamefont{Melchert}}, \bibnamefont{and}
  \bibinfo{author}{\bibfnamefont{A.~K.} \bibnamefont{Hartmann}},
  \bibinfo{journal}{Phys. Rev. E} \textbf{\bibinfo{volume}{79}},
  \bibinfo{pages}{031103} (\bibinfo{year}{2009}).

\bibitem[{\citenamefont{Melchert et~al.}(2010)\citenamefont{Melchert, Apolo,
  and Hartmann}}]{Oli10}
\bibinfo{author}{\bibfnamefont{O.}~\bibnamefont{Melchert}},
  \bibinfo{author}{\bibfnamefont{L.}~\bibnamefont{Apolo}}, \bibnamefont{and}
  \bibinfo{author}{\bibfnamefont{A.~K.} \bibnamefont{Hartmann}},
  \bibinfo{journal}{Phys. Rev. E} \textbf{\bibinfo{volume}{81}},
  \bibinfo{pages}{051108} (\bibinfo{year}{2010}).

\bibitem[{\citenamefont{Melchert and Hartmann}(2011{\natexlab{b}})}]{Oli11}
\bibinfo{author}{\bibfnamefont{O.}~\bibnamefont{Melchert}} \bibnamefont{and}
  \bibinfo{author}{\bibfnamefont{A.~K.} \bibnamefont{Hartmann}},
  \bibinfo{journal}{Eur. Phys. J. B} \textbf{\bibinfo{volume}{80}},
  \bibinfo{pages}{155} (\bibinfo{year}{2011}{\natexlab{b}}).

\bibitem[{\citenamefont{Melchert et~al.}(2011)\citenamefont{Melchert, Hartmann,
  and M\'ezard}}]{Oli11meanField}
\bibinfo{author}{\bibfnamefont{O.}~\bibnamefont{Melchert}},
  \bibinfo{author}{\bibfnamefont{A.~K.} \bibnamefont{Hartmann}},
  \bibnamefont{and} \bibinfo{author}{\bibfnamefont{M.}~\bibnamefont{M\'ezard}},
  \bibinfo{journal}{Phys. Rev. E} \textbf{\bibinfo{volume}{84}},
  \bibinfo{pages}{041106} (\bibinfo{year}{2011}).

\bibitem[{\citenamefont{Schramm}(2000)}]{schramm2000}
\bibinfo{author}{\bibfnamefont{O.}~\bibnamefont{Schramm}},
  \bibinfo{journal}{Israel J. Math.} \textbf{\bibinfo{volume}{118}},
  \bibinfo{pages}{221} (\bibinfo{year}{2000}).

\bibitem[{\citenamefont{Bernard et~al.}(2006)\citenamefont{Bernard, Boffetta,
  Celani, and Falkovich}}]{bernard2006}
\bibinfo{author}{\bibfnamefont{D.}~\bibnamefont{Bernard}},
  \bibinfo{author}{\bibfnamefont{G.}~\bibnamefont{Boffetta}},
  \bibinfo{author}{\bibfnamefont{A.}~\bibnamefont{Celani}}, \bibnamefont{and}
  \bibinfo{author}{\bibfnamefont{G.}~\bibnamefont{Falkovich}},
  \bibinfo{journal}{Nature Physics} \textbf{\bibinfo{volume}{2}},
  \bibinfo{pages}{124} (\bibinfo{year}{2006}).

\bibitem[{\citenamefont{Smirnov}(2006)}]{smirnov2006}
\bibinfo{author}{\bibfnamefont{S.}~\bibnamefont{Smirnov}},
  \bibinfo{journal}{Eur. Math. Soc.} \textbf{\bibinfo{volume}{2}},
  \bibinfo{pages}{1421} (\bibinfo{year}{2006}).

\bibitem[{\citenamefont{Smirnov}(2001)}]{smirnov2001}
\bibinfo{author}{\bibfnamefont{S.}~\bibnamefont{Smirnov}}, \bibinfo{journal}{C.
  R., Math., Acad. Sci. Paris} \textbf{\bibinfo{volume}{333}},
  \bibinfo{pages}{239} (\bibinfo{year}{2001}).

\bibitem[{\citenamefont{Amoruso et~al.}(2006)\citenamefont{Amoruso, Hartmann,
  Hastings, and A.}}]{amoruso2006}
\bibinfo{author}{\bibfnamefont{C.}~\bibnamefont{Amoruso}},
  \bibinfo{author}{\bibfnamefont{A.~K.} \bibnamefont{Hartmann}},
  \bibinfo{author}{\bibfnamefont{M.~B.} \bibnamefont{Hastings}},
  \bibnamefont{and} \bibinfo{author}{\bibfnamefont{M.~M.} \bibnamefont{A.}},
  \bibinfo{journal}{Phys. Rev. Lett.} \textbf{\bibinfo{volume}{97}},
  \bibinfo{pages}{267202} (\bibinfo{year}{2006}).

\bibitem[{pap()}]{papercore}
\bibinfo{note}{\textit{Papercore} is a free and open access database for
  summaries of scientific (currently mainly physics) papers, URL
  \url{http://www.papercore.org/}.}

\bibitem[{\citenamefont{Ahuja et~al.}(1993)\citenamefont{Ahuja, Magnanti, and
  Orlin}}]{ahuja1993}
\bibinfo{author}{\bibfnamefont{R.~K.} \bibnamefont{Ahuja}},
  \bibinfo{author}{\bibfnamefont{T.~L.} \bibnamefont{Magnanti}},
  \bibnamefont{and} \bibinfo{author}{\bibfnamefont{J.~B.} \bibnamefont{Orlin}},
  \emph{\bibinfo{title}{{Network Flows: Theory, Algorithms, and Applications}}}
  (\bibinfo{publisher}{Prentice Hall}, \bibinfo{year}{1993}).

\bibitem[{\citenamefont{Cook and Rohe}(1999)}]{cook1999}
\bibinfo{author}{\bibfnamefont{W.}~\bibnamefont{Cook}} \bibnamefont{and}
  \bibinfo{author}{\bibfnamefont{A.}~\bibnamefont{Rohe}},
  \bibinfo{journal}{INFORMS J. Computing} \textbf{\bibinfo{volume}{11}},
  \bibinfo{pages}{138} (\bibinfo{year}{1999}).

\bibitem[{\citenamefont{Hartmann and Rieger}(2001)}]{opt-phys2001}
\bibinfo{author}{\bibfnamefont{A.~K.} \bibnamefont{Hartmann}} \bibnamefont{and}
  \bibinfo{author}{\bibfnamefont{H.}~\bibnamefont{Rieger}},
  \emph{\bibinfo{title}{{Optimization Algorithms in Physics}}}
  (\bibinfo{publisher}{Wiley-VCH}, \bibinfo{address}{Weinheim},
  \bibinfo{year}{2001}).

\bibitem[{com()}]{comment_cookrohe}
\bibinfo{note}{For the calculation of minimum-weighted perfect matchings we use
  Cook and Rohes blossom4 extension to the Concorde library.},
  \urlprefix\url{http://www2.isye.gatech.edu/~wcook/blossom4/}.

\bibitem[{\citenamefont{{Beffara}}(2004)}]{beffara2004}
\bibinfo{author}{\bibfnamefont{V.}~\bibnamefont{{Beffara}}},
  \bibinfo{journal}{Ann. Probab.} \textbf{\bibinfo{volume}{32}},
  \bibinfo{pages}{2606} (\bibinfo{year}{2004}).

\bibitem[{\citenamefont{{Beffara}}(2008)}]{beffara2008}
\bibinfo{author}{\bibfnamefont{V.}~\bibnamefont{{Beffara}}},
  \bibinfo{journal}{Ann. Probab.} \textbf{\bibinfo{volume}{36}},
  \bibinfo{pages}{1421} (\bibinfo{year}{2008}).

\bibitem[{\citenamefont{{Buldyrev} et~al.}(2004)\citenamefont{{Buldyrev},
  {Havlin}, {L\'{o}pez}, and {Stanley}}}]{havlin2004}
\bibinfo{author}{\bibfnamefont{S.~V.} \bibnamefont{{Buldyrev}}},
  \bibinfo{author}{\bibfnamefont{S.}~\bibnamefont{{Havlin}}},
  \bibinfo{author}{\bibfnamefont{E.}~\bibnamefont{{L\'{o}pez}}},
  \bibnamefont{and} \bibinfo{author}{\bibfnamefont{H.~E.}
  \bibnamefont{{Stanley}}}, \bibinfo{journal}{Phys. Rev. E}
  \textbf{\bibinfo{volume}{70}}, \bibinfo{pages}{035102(R)}
  (\bibinfo{year}{2004}).

\bibitem[{\citenamefont{Risau-Gusman and Rom\'{a}}(2008)}]{roma2008}
\bibinfo{author}{\bibfnamefont{S.}~\bibnamefont{Risau-Gusman}}
  \bibnamefont{and} \bibinfo{author}{\bibfnamefont{F.}~\bibnamefont{Rom\'{a}}},
  \bibinfo{journal}{Phys. Rev. B} \textbf{\bibinfo{volume}{77}},
  \bibinfo{pages}{134435} (\bibinfo{year}{2008}).

\bibitem[{\citenamefont{Picco and Santachiara}(2008)}]{picco2008}
\bibinfo{author}{\bibfnamefont{M.}~\bibnamefont{Picco}} \bibnamefont{and}
  \bibinfo{author}{\bibfnamefont{R.}~\bibnamefont{Santachiara}},
  \bibinfo{journal}{Phys. Rev. Lett.} \textbf{\bibinfo{volume}{100}},
  \bibinfo{pages}{015704} (\bibinfo{year}{2008}).

\bibitem[{\citenamefont{Chatelain}(2010)}]{chatelain2010}
\bibinfo{author}{\bibfnamefont{C.}~\bibnamefont{Chatelain}},
  \bibinfo{journal}{J. Stat. Mech.} p. \bibinfo{pages}{P08004}
  (\bibinfo{year}{2010}).

\bibitem[{\citenamefont{Bernard et~al.}(2007)\citenamefont{Bernard, Le~Doussal,
  and Middleton}}]{bernard2007}
\bibinfo{author}{\bibfnamefont{D.}~\bibnamefont{Bernard}},
  \bibinfo{author}{\bibfnamefont{P.}~\bibnamefont{Le~Doussal}},
  \bibnamefont{and} \bibinfo{author}{\bibfnamefont{A.~A.}
  \bibnamefont{Middleton}}, \bibinfo{journal}{Phys. Rev. B}
  \textbf{\bibinfo{volume}{76}}, \bibinfo{pages}{020403(R)}
  (\bibinfo{year}{2007}).

\bibitem[{\citenamefont{{Schramm}}(2001)}]{schramm2001}
\bibinfo{author}{\bibfnamefont{O.}~\bibnamefont{{Schramm}}},
  \bibinfo{journal}{Elect. Comm. in Probab.} \textbf{\bibinfo{volume}{6}},
  \bibinfo{pages}{115} (\bibinfo{year}{2001}).

\bibitem[{\citenamefont{{Gamsa} and {Cardy}}(2007)}]{cardy2007}
\bibinfo{author}{\bibfnamefont{A.}~\bibnamefont{{Gamsa}}} \bibnamefont{and}
  \bibinfo{author}{\bibfnamefont{J.}~\bibnamefont{{Cardy}}},
  \bibinfo{journal}{J. Stat. Mech.} p. \bibinfo{pages}{P08020}
  (\bibinfo{year}{2007}).

\bibitem[{\citenamefont{Cayley}(1846)}]{cayley1846}
\bibinfo{author}{\bibfnamefont{A.}~\bibnamefont{Cayley}},
  \bibinfo{journal}{Crelle's J.} \textbf{\bibinfo{volume}{\textbf{32}}},
  \bibinfo{pages}{119} (\bibinfo{year}{1846}).

\bibitem[{\citenamefont{Hartmann}(2009)}]{practicalGuide2009}
\bibinfo{author}{\bibfnamefont{A.~K.} \bibnamefont{Hartmann}},
  \emph{\bibinfo{title}{{Practical Guide to Computer Simulations}}}
  (\bibinfo{publisher}{World Scientific}, \bibinfo{address}{Singapore},
  \bibinfo{year}{2009}).

\end{thebibliography}

\end{document}